\documentclass[twocolumn,amsmath,amssymb,aps,pra,floatfix,nofootinbib,superscriptaddress]{revtex4-1}
\usepackage{graphicx}
\usepackage{bm}

\begin{document}

\title{Towards non-invasive cancer diagnostics and treatment based on electromagnetic fields, optomechanics and microtubules}


\author{V. Salari}
\email{vahidsalari@cc.iut.ac.ir}
\affiliation{Department of Physics, Isfahan University of Technology, Isfahan 84156-83111, Iran}
\author{Sh. Barzanjeh}
\email{shabir.barzanjeh@ist.ac.at}
\affiliation{Institute of Science and Technology~(IST) Austria, 3400 Klosterneuburg, Austria}
\author{M. Cifra}
\affiliation{Institute of Photonics and Electronics, The Czech Academy of Sciences, Chabersk\'{a} 57, 182 00 Prague, Czech Republic}
\author{C. Simon}
\affiliation{Department of Physics and Astronomy, University of Calgary, Calgary T2N 1N4, Alberta, Canada}
\affiliation{Institute for Quantum Science and Technology, University of Calgary, Calgary T2N 1N4, Alberta, Canada}
\author{F. Scholkmann}
\affiliation{Biomedical Optics Research Laboratory, Department of Neonatology, University Hospital Zurich, University of Zurich, CH-8091 Zurich, Switzerland}
\affiliation{Research Office for Complex Physical and Biological Systems (ROCoS), CH-8038 Zurich, Switzerland}
\author{Z. Alirezaei}
\affiliation{Department of Medical Physics, Isfahan University of Medical Sciences, Isfahan, Iran}
\author{J. A. Tuszynski}
\affiliation{Department of Oncology, University of Alberta, Cross Cancer Institute, Edmonton T6G 1Z2, Alberta, Canada}
\affiliation{Department of Physics, University of Alberta, Edmonton,  AB T6G 2E1, Canada}

\date{\today}

\begin{abstract}

In this paper, we discuss biological effects of electromagnetic (EM) fields in the context of cancer biology. In particular, we review the nanomechanical properties of microtubules (MTs), the latter being one of the most successful targets for cancer therapy. We propose an investigation on the coupling of electromagnetic radiation to mechanical vibrations of MTs as an important basis for biological and medical applications. In our opinion optomechanical methods can accurately monitor and control the mechanical properties of isolated MTs in a liquid environment. Consequently, studying nanomechanical properties of MTs may give useful information for future applications to diagnostic and therapeutic technologies involving non-invasive externally applied physical fields. For example, electromagnetic fields or high intensity ultrasound can be used therapeutically avoiding harmful side effects of chemotherapeutic agents or classical radiation therapy.
\end{abstract}

\maketitle

\section{Introduction}
In spite of many efforts and important advances in cancer diagnostics and treatment, there are still millions of people dying each year throughout the world due to cancer, and the ”war on cancer” that was declared by the US President Richard Nixon in 1971 is still far from being won 46 years later. Conversely, the cancer incidence will most probably increase further \cite{Smith_2009, Jemal_2010} while the progress in cancer treatments has been stagnant for decades \cite{Soto, hanahan2014rethinking} with the exception of recent introduction of immmunotherapies. We should mention, however, that improvements in diagnostic imaging, especially using magnetic resonance imaging (MRI), and novel surgery approaches have led to progress in the field of oncology with improvements in clinical outcomes accruing gradually over the past few decades. The new surgery techniques for example enable a better delineation of the tumour area and the surrounding structures, allowing the medical oncologists to reduce the tumor margin and hence decrease damage to the normal tissue \cite{Alpuente} leading to improved quality of life of the patentients. However, truly cutting-edge technologies are still to be discovered and implemented within the arsenal of cancer therapy modalities. In particular, the roles of biochemical signaling pathways, and biophysical aspects (nanomechanical states of the cellular- and subcellular structures, endogenous bioelectric current/potential and electromagnetic field) have not been fully understood and exploited in the field of oncology \cite{Levin, suresh2007biomechanics}.\\
The aim of the present review article is to outline why it is worthwhile to focus on biophysical aspects with regard to future diagnosis and treatment approaches for cancer. In particular, we discuss why the bioelectric and mechanical properties of microtubules might play a significant role in cancer, whose better understanding could lead to the development of novel therapies. Biolectricity is a basic phenomenon associated with cellular and subcellular structures \cite{funk2015endogenous, levin2014molecular, levin2012molecular}. Most of the subcellular biomolecules (e.g. DNA, RNA, tubulin, actin, septin, etc.) are either charged and hence surrounded by counter-ions or endowed with high electric dipole moments that can enagage in dipole-dipole interactions and polarize electrically their local environment. Living organisms are replete with both moving and oscillating electric charges and can thus be regarded as complex electrochemical and mechanical systems. Complex patterns of direct current (DC) electric fields present within living organisms are key factors in morphogenesis and contain part of the information needed to produce a three-dimensional organism \cite{Northrop}. These factors, in our opinion, need to be addressed when finding novel methods of cancer diagnosis and treatment.

\section{Electromagnetic fields affect cancer cells}

It has been shown that extremely low-frequency (ELF), pulsed electromagnetic fields (PEMF) and sinusoidal electromagnetic fields (SEMF) can induce tumor cell apoptosis, inhibit angiogenesis, impede proliferation of neoplastic cells, and cause necrosis non-invasively, whereas human lymphocytes are negligibly affected \cite{Berg}. Some studies describe the effects of intense ($>$ 0.1 MV/m) nanosecond (10-300 ns) pulsed electric fields on mammalian cell structure and function. As the pulse durations decrease, effects on the plasma membrane decrease and effects on intracellular signal transduction mechanisms increase \cite{Beebe}. 

\textit{Low-power} EMF within the range of 0.1-40 MHz may impair the DNA strand and cause inhibition of proliferation of the gallbladder cancer cells, and these effects are related to the frequency of the electromagnetic fields but not in a linear fashion \cite{Chen}. Kirson et al. \cite{Kirson1} have recently demonstrated that 100 KHz to 1 MHz AC fields have significant specific effects on dividing cells. The basis of these effects during cytokinesis was hypothesized to be the unidirectional dielectrophoretic forces induced by the inhomogeneous fields at the cleavage furrow separating the daughter cells that interfere with the orientation of spindle microtubules \cite{Kirson2}. A review of other possible mechanisms involved in the interactions of these fields, dubbed TTFields (Tumor Treating Fields) and cancer cells has been recently published \cite{Tuszynski}. It is worth noting that in addtion to microtubules, actin filaments, DNA and even ion channels may be affected by TTFields and investigations into specific molecular mechanisms are on-going. As an additional electromagnetic mechanism with potential cancer treatment application, a study was undertaken to examine whether millimeter electromagnetic waves (MMWs) irradiation (42.2 GHz) can inhibit tumor metastasis enhanced by cyclophosphamide (CPA), an anticancer drug. \cite{Logani}.

Concerning \textit{high-power} EMF, for example Elson (2009) focused on the potential of strong magnetic fields to play a role in cancer treatment. Results of this study show that pulsed magnetic field (PMF) in combination with ultraviolet C (UVC) have the ability to augment the cell killing effects of UVC radiation. In addition, the effects appear to be greater when PMF and UVC are applied at the same time \cite{Elson}.
Mitochondria are well known to play an important role in apoptosis. Steep pulsed electric fields (SPEF) could induce apoptosis markedly (P-value less than 0.01); SPEF with lower voltage (200V) and longer width (1.3 $\mu s$) could induce apoptosis more effectively than SPEF with higher voltage (600V) and shorter width (100ns). These experimental results provide a possible mechanism and parameter selection basis for tumor treatment using SPEF \cite{Mi}. There are many reports of enhanced transcription and replication in different cell culture systems exposed to electromagnetic fields, and reports of cytoreduction (necrosis and apoptosis) in tumors transplanted into animals exposed to similar, often much stronger electromagnetic fields, but where heating is negligible. Although the mechanism of inducing apoptosis has not been characterized yet, one major candidate for the initiation of such a process is the production of numerous breaks in DNA, and the inhibition of DNA repair processes, leading to the initiation of the apoptotic (programmed cell death) process \cite{Elson}. 

Interestingly, electromagnetic frequencies at which cancer cells become sensitive appear to be tumor-specific and hence treatment with tumor-specific frequencies is feasible, well tolerated and may have biological efficacy in patients with advanced cancer. A study that examined a total of 163 patients diagnosed with various types of cancer has identified a total of 1524 distinct frequencies ranging from 0.1 Hz to 114 kHz. Most frequencies (57 to 92 percent) were specific for a single tumor. 
These observations suggest that electromagnetic fields, which are amplitude-modulated at tumor-specific frequencies, do not act solely on tumors but may have wide-ranging effects on tumor-host interactions, e.g. immune modulation \cite{Barbault}. 

\textit{In vitro }effects of electromagnetic fields appear to be related to the type of electromagnetic field applied. It has been shown that human osteoblasts display effects of BEMER (Bio-Electro Magnetic Energy Regulation) type electromagnetic field (BTEMF) on gene regulation. Effects of BTEMF on gene expression in human mesenchymal stem cells and chondrocytes have been analyzed. Results indicate that BTEMF in human mesenchymal stem cells and chondrocytes provide the first indications to understanding therapeutic effects achieved with BTEMF stimulation \cite{Walther}.

Phenotypic changes in human breast cancer cells following low-level magnetic field (MF) exposure previously reported. Proteomic methods were used to investigate the biochemical effect of MF exposure in SF767 human glioma cells. Protein alterations were studied after exposure to 1.2 microTesla (microT) MF [12 milliGauss (mG), 60 Hertz (Hz)]. The results suggest that the analysis of differentially expressed proteins in SF767 cells may be useful as biomarkers for biological changes caused by exposure to magnetic fields \cite{Kanitz}.

Qutob et al. (2006) showed that there was no evidence that non-thermal RF fields can affect gene expression in cultured U87MG glioblastoma cells relative to the non-irradiated control groups, whereas exposure to heat shock at 43 degrees C for 1 h up-regulated a number of typical stress-responsive genes in the positive control group \cite{Qutob}.
Gap junction genes are recognized as tumor suppressors \cite{Omori, Mesnil, Yamasaki} and effects on gap junctional communication also provide an appealing model for explaining tumor growth induced by exposure to weak magnetic fields. ELF exposure generally does not transmit nearly enough energy to cause mutagenesis of DNA, but has been shown to affect gap junction states and thus potentially to control proliferation and differentiation \cite{Hu, Yamaguchi, Huang}.

In the MHz region, several studies investigated the effect of the application of electromagnetic field in the MHz for cancer treatment. Normally, the intensity applied was high, inducing thermal effects. That the choice of the specific modulation of the EMF is important was shown by Andocs et al. \cite{andocs2009strong}, demonstrating the the application of modulated EMF (13.56 MHz) causes a synergistically anticancer effect due to a hermal and a non-thermal mechanism triggered (modulated electrohyperthermia). Subsequent work showed that this treatment causes DNA-fragmentation \cite{meggyeshazi2014dna} and up-regulation of heat-shock proteins \cite{andocs2015upregulation} in the cancer cells. The superiority of using 13.56 MHz modulated electrohyperthermia in comparison to only classical hyperthermia to treat cancer was demonstrated recently \cite{andocs2016comparison}.

In the GHz range, most of the studies used strong GHZ EMF for treating tumors by induce thermal effects \cite{converse2006computational, nguyen2017three, mendecki1978microwave}. However, EMF in the GHz range can also act as a co-cancerogen \cite{lerchl2015tumor, hardell2015increasing}, making the application of GHz EMF for cancer treatment not as an optimal approach.


Terahertz (THz) radiation occupies a broad band of the EM spectrum between microwave and infrared frequencies, and is therefore non-ionizing. The THz region covers the frequency range from 0.1 to 10 THz and it offers non-invasive diagnostic capabilities that fill the “gaps between x-rays, MRI, and the isible range \cite{Savage}. Applications of THz waves to the diagnosis of melanoma \cite{fitzgerald}, basal cell carcinoma \cite{woodward}, and breast cancer \cite{ashworth, fitzgerald2006, chen2011}  have already been demonstrated. THz technology has led to the development of commercially available diagnostic medical applications such as THz Pulsed Spectroscopy \cite{ashworth} and THz Pulsed Imaging \cite{fitzgerald2006, wilmink}, which offer excellent contrast between diseased and healthy tissues. The first clinical trials of THz imaging as an intra-operative tool during cancer surgery are underway \cite{Titova2013}. Studies on stem cells suggest that exposure to broad-spectrum THz pulses affects cell differentiation and gene expression \cite{bock, alexandrov2013} at both the transcript and protein levels. The mechanism by which THz radiation interacts with biological systems is fundamentally different to that of conventional ionizing therapies, due to its resonant effects on cell membranes \cite{doria}, proteins \cite{falconer} and nucleic acids \cite{alexandrov2010, fischerr,fischer2005}. It has been recently demonstrated that intense, picosecond THz pulses induce changes in cellular functions\cite{titova, titova2013intense,titova3, titova4}. Exposure of human tissue to intense THz pulses was found to activate the DNA damage response (DDR), and affect expression levels of many proteins, especially cell-cycle regulatory proteins offering a potential for therapeutic applications of this novel modality. In addition, a combination of this modality can be considered with standard chemotherapy since sub-$\mu$s pulsed electric fields applied to tumours through electrodes have been shown to permeabilize tumour cell membranes to cytotoxic agents\cite{lucas, kubota, gothelf}. Consequently, intense THz pulses may significantly lower the required therapeutic doses of cytotoxic drugs. However, the fundamental mechanisms of interaction of THz radiation with biological systems so far remain elusive. 

\section{Microtubules} 
It is well known that MTs, microfilaments and intermediate filaments are the main components of cytoskeleton of eukaryotic cells. MTs are the most rigid protein polymer among the three types of cytoskeletal filaments, which form the architecture of the cell. They exhibit unique physical behaviour and form special structures well suited for their own cellular functions \cite{Alberts}. The structure of MTs is cylindrical, and it typically involves 13 parallel protofilaments, which are connected laterally into hollow tubes. MTs have 25 nm external and 15 nm internal diameters. The length of MTs can vary from tens of nanometers to hundreds of microns \cite{Civalek}. MT biological functions rely on two essential properties. First, they are dynamic polymers that are assembled and disassembled rapidly in a fashion coordinated with motile reactions; second, they are relatively rigid structures able to resist the pico-Newton level forces exerted by kinesin and dynein motor proteins, and they provide the required mechanical stiffness for cilia and flagella \cite{Venier}. 

Mechanical properties of MTs largely determine their functions. Quantifying the way they resist mechanical deformation by determining their Young's and shear modulus can lead to a better understanding of all the vital physiological mechanisms in which MTs are involved. For instance, it would be favorable for the stable MTs of the axon to be stiff and straight to support the extended structure required for long-distance axonal transport. Conversely, MTs in a proliferating cell should be dynamic and flexible to enable rapid redistribution during transitions between interphase and mitosis \cite{Hawkins}. 

However, measuring and understanding MTs' mechanical properties is not a simple task. Two decades of measurements involving different techniques such as optical tweezers \cite{Kurachi}, hydrodynamic flow \cite{Venier}, atomic force microscope (AFM) \cite{Vinckier, Kis}, and persistence length observations \cite{Mizushima}, resulted in values of elastic (shear and Young's) modulus spanning a range of values between 1 MPa and 7 GPa \cite{Jack1}. For instance, experiments involving the MTs with lengths 24-68 nm yielding a value of 2 GPa for MTs assembled from pure bovine-brain tubulin \cite{Kasas}.
 Short MTs are flexible due to a low value of the shear modulus while longer tubes become more rigid, which is when the Young's modulus dominates the mechanical behaviour. Measurements on longer MTs would therefore provide better estimates of the Young's modulus, because neglecting the influence of shearing would introduce a smaller error.

Since microtubules in biological conditions are often subject to a dynamic load, vibration analysis suggests itself as a method to study their dynamic response. Vibration normal modes describe the preferential pattern of structural dynamics of a microtubule, whereby its response to a time-varying force can be represented by a combination of these vibration modes. In addition, there are several hypotheses that ascribe biological relevance to the vibrations themselves. Furthermore, microtubule vibration mode patterns are reminiscent to buckling pattern, so the underlying mathematical apparatus is similar \cite{Civalek}.

There are generally two modeling approaches that have been developed to study microtubule mechanics and vibrations. A continuum mechanics model \cite{Sirenko, portet2005elastic, farajpour2014surface, arani2013nonlinear, mustapha2016torsional, Wang, wang2009dynamic, daneshmand2012coupled} and a discrete model \cite{pokorny1997vibrations, deriu2010anisotropic, havelka2017deformation}. The gap between high-precision molecular modeling and continuum modeling can be bridged by an atomistic-continuum model; the first of such models to study the topic herein was presented by Liew et al. \cite{Xiang, liew2015mechanical}.

These theoretical treatments of the microtubule structure disclose their vibrational normal modes in a wide frequency range from acoustic to GHz frequencies \cite{Sirenko, Wang, Qian, Jack1, kucera2016, havelka2017deformation}. For instance It has been shown that the microtubule lengths L in terms of their fundamental bending mechanical resonance frequencies between 100 and 200 kHz in vitro (20~$^\circ$C) \cite{Dubost}. Elastic wave propagation in MTs was analyzed in different works \cite{Sirenko, Wang, Qian}. Dependence of frequency or velocity of propagation on the wave vector was evaluated for isotropic and orthotropic shell models with different parameters. In particular, the resonant condition for a 10 nm long MT corresponding to half of the wavelength at a frequency of about 460 MHz may occur for longitudinal oscillations with Young's modulus 1.7-2 GPa. These results hold for the orthotropic microtubule axisymmetric (n = 0) and nonaxisymmetric (n = 1) shell models \cite{Pokorny, Sirenko, Wang}. Numerical calculations based on recently obtained experimental data for Young's modulus of MT, show that MT-water system supports interface elastic waves with maximal frequencies in a GHz range. In fact, \cite{Sirenko} performed theoretical analysis for elastic vibrations of MT immersed in water and found that this system supports nonradiative elastic waves localised in the vicinity of the MT wall with maximal frequencies of order of tens of GHz. In the long wavelength limit, there exist three axisymmetric acoustic waves with propagation speed of approximately 200-600 m/s and an infinite set of helical waves with a parabolic dispersion law \cite{Sirenko}.

The role of mechanical vibrations of MT is not known in biology so far \cite{kucera2016}. The fundamental issue for any biological relevance of MT vibrations or further phenomena assuming MT vibrations is the damping of MT vibrations. It is generally considered that protein and MT normal mode vibrations are overdamped \cite{howard2001}. Some works estimate that, depending on the type of the vibration mode and lowered coupling with the MT viscous environment, the quality factor Q of the vibration modes may be in the range of 0.01 - 10 \cite{kucera2016}, hence reaching to underdamped regime. However, there are no solid experimental data on the damping of MT vibrations so far, only extensive theoretical works. 

Tubulin, a MT subunit, is a protein which has rather high charge and dipole moment compared to most other proteins \cite{tuszynski2003evolution}. Hence, it is natural to suggest that MT vibrations, if underdamped and excited, will be accompanied by an electrical field of the same frequency as was originally proposed by Pokorn\'{y} et al. \cite{pokorny1998electric}. Several recent works developed this idea \cite{cifra_electric_2010, kucera_mechano-electrical_2012} including electromechanical vibrational models of whole cell microtubule network \cite{havelka_high-frequency_2011, havelka_electro-acoustic_2014} and multi-mode vibration of single microtubule \cite{havelka_multi-mode_2014}.
Charge and dipole moment of MT as well as of tubulin, other proteins and polar nanoobjects in general is also a key to coupling external electromagnetic field to vibrations of such objects; the coupling is significant only when vibrations are sufficiently underdamped \cite{krivosudsky2016microwave}. Single protein normal vibration modes are in the range of cca. 0.03--3 THz \cite{karplus1981internal, turton2014terahertz, wheaton2015probing} together with MT vibration band (kHz - GHz) can hypothetically enable interaction with electromagnetic field at frequencies across many orders of magnitude.\\
It was hypothesized that vibrations of MT cytoskeleton generate coherent electromagnetic field which plays role in organization of processes in living cells and that this field is perturbed in cancer \cite{pokorny2013postulates}. Within this hypothesis it is considered that the damping of MT vibrations is caused by the ambient medium, i.e., by the cytosol water. It is proposed that in cancer the changes in mitochondrial metabolism lead to change of the water structure around MT and to increased damping of Mt vibrations might cause a shift in the resonance frequency of oscillations in cells. Frequency changes in cancer cells were also predicted by Fr\"{o}hlich \cite{Froehlich}. A peculiar cancer diagnostic method developed by Vedruccio \cite{Vedruccio} claimed to exploit frequency selective effects of the interaction of the external electromagnetic field with cancer cells was interpreted using this hypothesis. Hypotheses of Pokorn\'{y} and Fr\"{o}hlich also inspired a number of experimental works aiming to directly electronically detect electromagnetic activity of living cells in radiofrequency and microwave bands \cite{holzel2001electric, jelinek2007measurement, jelinek_measurement_2009}. However, solid evidence for such cellular electromagnetic activity remains elusive \cite{kucera_spectral_2015}. However, unless the possibility of underdamped MT vibrations and endogenous excitation is proved, these hypotheses of highly coherent biological electromagnetic field remain unrealistic. One important piece of puzzle could be brought by elucidation of one of the crucial assumptions in these hypotheses: low damping of microtubule vibrations. Knowledge of dynamic mechanical properties is also essential to assess effects of ns intense electric pulses which have been demonstrated to affect cytoskeleton \cite{carr2017calcium}, thus opening a new avenue how to disrupt cell divison with potential cancer applications. However, exact mechanisms of action remains unclear. 
Thus, to enable new perspective diagnostic and therapeutic methods based on MT monitoring and manipulation, a rigorous experimental analysis of microtubule vibrations and dynamic mechanical properties is needed.

\section{Monitoring mechanical vibrations of microtubules via optomechanical coupling}
The emerging field of optomechanics is concerned with the study of the mechanical effects of light on mesoscopic and macroscopic mechanical oscillators. These phenomena have been realized in optomechanical systems consisting of  an optical cavity with a movable end-mirror or with a membrane-in-the middle. The radiation pressure exerted by the light inside the optical cavity couples the moving mirror or the membrane which acts as a mechanical oscillator to the optical field. This optomechanical coupling has been employed for a wide range of applications such as the cavity cooling of microlevers and nanomechanical resonators to their quantum mechanical ground state \cite{Genes2008, Teufel1, Liberato, Ludwig, Barzanjeh2}, producing high precision detectors for measuring weak forces and small displacements and also for fundamental studies of the transition between the quantum and the classical world \cite{Bradaschia,LaHaye,Kippenberg,O'Connell,Teufel1,Poot,Pikovski,Bawaj,Liberati2,Barzanjehdis}.

Optomechanical systems can also be applied for the sensitive detection of physical quantities such as spin~\cite{spin1,spin2}, atomic/molecular mass~\cite{mol1,mol2,mol3}, the concentration of biologically relevant molecules~\cite{Shekhawat}, and thermal fluctuations~\cite{Badzey,Paul,Tamayo}, as well as for frequency conversion~\cite{Barzanjeh2011,Barzanjeh1,Andrews,Barzanjehconf,Barzanjeh2015}. Nanomechanical resonators~(NMRs) with resonance frequencies in the GHz regime can be now fabricated~\cite{O'Connell,Huang,Peng,Painter1} and they are suitable candidates for the study of the quantum behavior at the mesoscopic scale~\cite{O'Connell,Painter1,Painter2}. These GHz NMRs are characterized by reduced dimensions and therefore by very low masses, and at the same time, in this regime the nonlinear behavior of the mechanical systems becomes more relevant, consequently offering interesting theoretical~\cite{Lifshitz,Cross,Katz, Barzanjehnon,Rips,Rips1,Shah,BarzanjehJJ} and experimental challenges~\cite{Yu,Aldridge,Kozinsky}. These high-frequency resonators operating in the nonlinear regime open up new possibilities for the realization of novel devices and applications of NMR and nanoelectromechanical systems~\cite{Lupascu,Woolley}.

\begin{figure}
\caption{Top view of the optomechanical setup for monitoring mechanical vibrations in microtubules}
\centering
\includegraphics[width=0.5\textwidth]{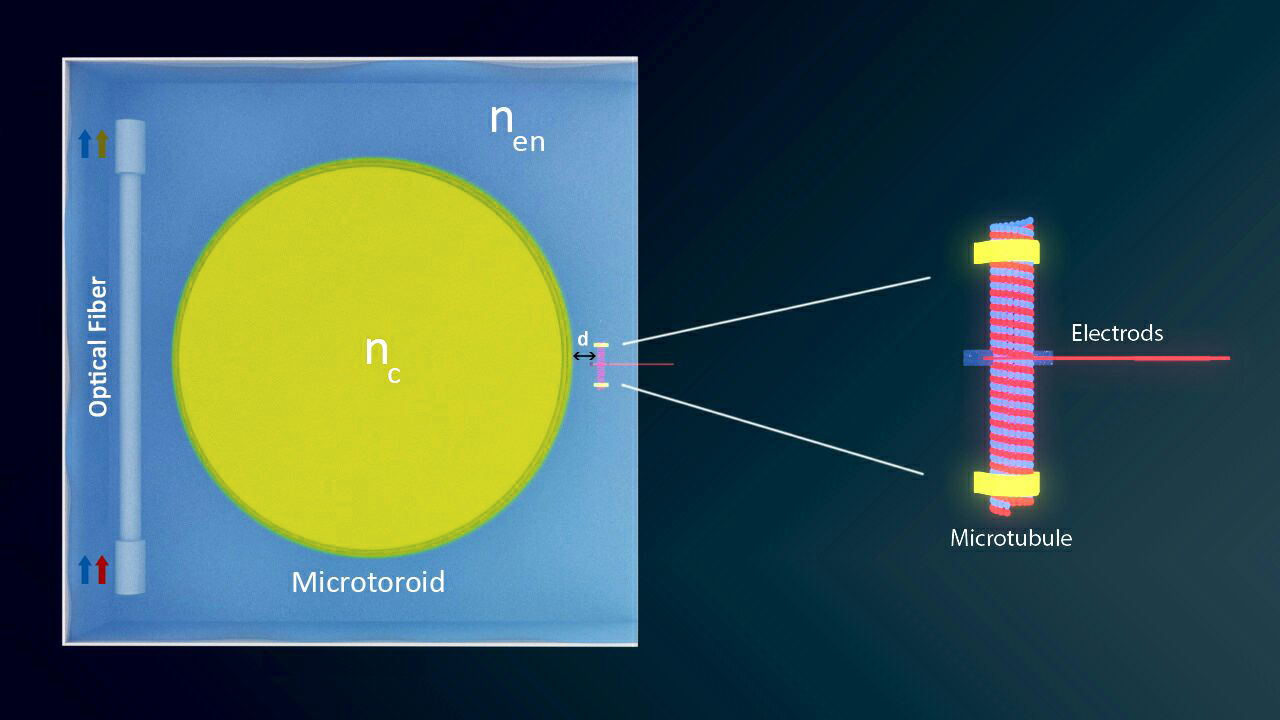}
\end{figure}

\begin{figure}
\caption{Side view of the optomechanical setup for monitoring mechanical vibrations in microtubules}
\centering
\includegraphics[width=0.5\textwidth]{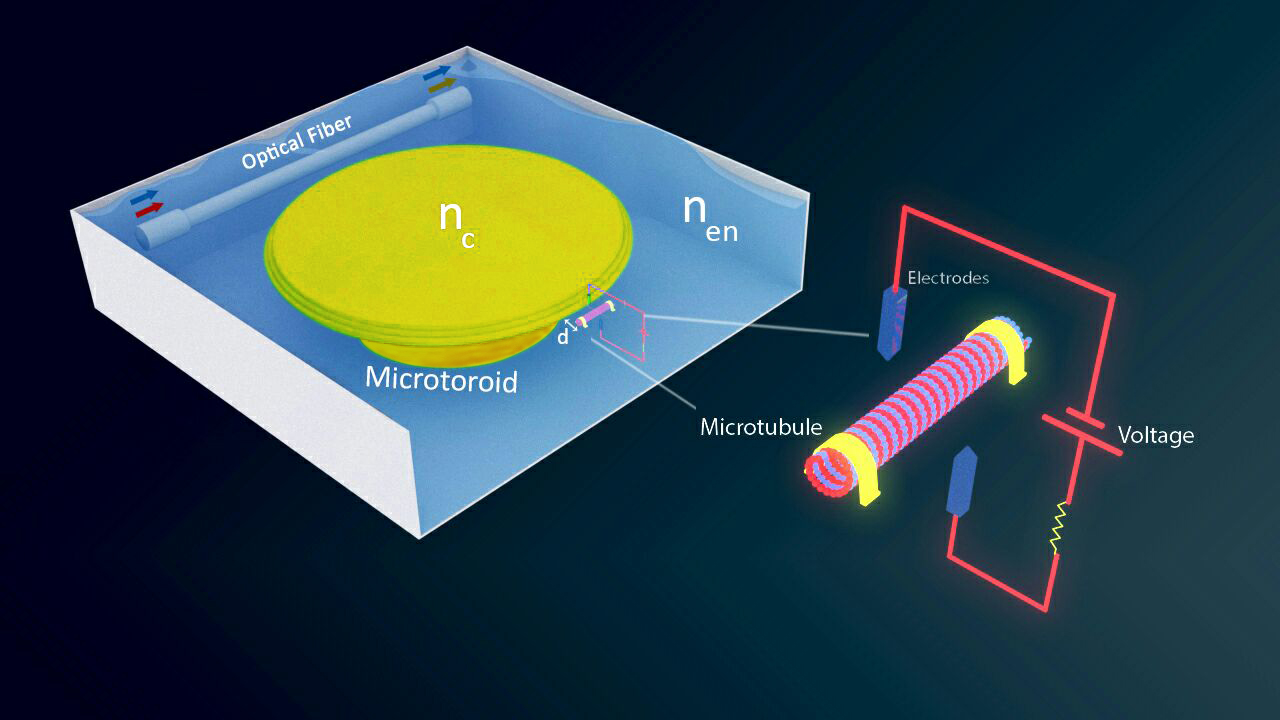}
\end{figure}


The optomechanical system can also be employed to observe the vibrations of isolated microtubules \cite{ShabirOPTO}. Information about the microtubule mechanical vibrations can be obtained by coupling the microtubule to an optical cavity. In fact, the optomechanical coupling between the microtubule and the optical field of the cavity modifies the response of the cavity field results in appearance of electromagnetically induced  transparency peaks in the transmission feature of the optical probe field. The center frequency and linewidth of the transparency peak give the resonance frequency and damping rate of the vibrational mode of the microtubule. By properly selecting the parameters of the system, one can observe up to 1GHz vibration for the microtubule. The dielectric properties of the microtubule, however, raises the possibility to control the vibration of the microtubule by positioning tip electrodes close to surface of the microtubule. 
Applying voltage on the electrode plates creates an effective external force on the microtubule, modifies the resonance frequency of the microtubule vibration \cite{ShabirOPTO}.

\section{Summary}
Microtubules, key structures forming the cellular skeleton, have been among the most successful targets for anticancer therapy. Any interference with their functioning, especially during mitosis, can control the replication of a cancer cell. However, chemotherapeutic techniques to disrupt microtubules have several side effects on healthy cells, making chemotherapy a less than ideal modality to suppress cancer proliferation.
It has been predicted that the mechanical properties of microtubules (such as their vibration frequencies) are different in cancer cells compared to healthy cells. Inspired by this, we believe that cancer treatment (and detection) may be possible based on the detection and control of microtubule mechanical vibrations in cells exposed to non-invasive external radiations (e.g. electromagnetic or ultrasound). We have proposed an optomechanical method to control and read out the vibrations of an isolated microtubule. This can help determine which frequencies can cause breakage in microtubules of cancer cells using resonance effects in microtubules from an external field. Moreover, this method may help to recognize cancer cells based on special frequencies in microtubules. So far, measuring a broad spectrum of mechanical vibrations of microtubules has not been an easy task, and there are only a small number of studies in this context. In order to improve our knowledge about the mechanical vibrations of microtubules, we proposed an optomechanical technique for measurement of microtubule dynamics at room temperature. Our approach is a step forward for monitoring mechanical frequencies of microtubules in a broad spectral range for a potential application in medical diagnosis and treatment. This may help scientists in the future to supplement standard cancer chemo- and radiotherapy approaches with non-ionizing physical fields to be used as therapeutic and possibly even diagnostic methods.

It is a great goal to reach simple and non-invasive methods for diagnosis and treatment of diseases like cancer. We have proposed that an optomechanical setup can help us to monitor the all possible vibrations in a MT {\it in vitro} for a potential application in cancer diagnosis and treatment. In fact, if we know what frequency can destruct the MT structure it can be  useful for cancer treatment since cancer cells need MTs for cell division and if they are disrupted or destroyed their growth can be stopped or tumor can be vanished. This type of disruption can be done via an external weak EM signal or an ultrasound signal which can be focused on the cancerous cells {\it in vivo} and their growth can be controlled by these external signals with special frequencies.  

\section{Acknowledgements}The work of SB has been supported by the European Union’s Horizon 2020 research and innovation program under the Marie Sklodowska Curie grant agreement No MSC-IF 707438 SUPEREOM. JAT gratefully acknowledges funding support from NSERC (Canada) for his research. MC acknowledges support from the Czech Science Foundation, projects no. 15-17102S and 17-11898S and he participates in COST Action BM1309, CA15211 and bilateral exchange project between Czech and Slovak Academies of Sciences, no. SAV-15-22.

\end{document}